\documentclass{scrartcl}
\pdfoutput=1
\usepackage[utf8]{luainputenc}
\usepackage[USenglish]{babel}
\usepackage{csquotes}

\usepackage[a4paper,top=27mm,bottom=25mm,inner=25mm,outer=20mm]{geometry}
\usepackage[plainpages=false,pdfpagelabels,hidelinks,unicode]{hyperref}

\usepackage[%
  backend=bibtex,bibencoding=ascii,
  style=numeric-comp,
  giveninits=true, uniquename=init, 
  natbib=true,
  url=true,
  doi=true,
  isbn=false,
  backref=false,
  maxnames=99,
  ]{biblatex}
\usepackage{siunitx}

\usepackage{graphicx}
\usepackage[small]{caption}
\usepackage{subcaption}

\begingroup\expandafter\expandafter\expandafter\endgroup
\expandafter\ifx\csname pdfsuppresswarningpagegroup\endcsname\relax
\else
\fi

\usepackage{ifluatex}
\ifluatex
  \usepackage[no-math]{fontspec}
\else
  \usepackage[T1]{fontenc}
\fi
\usepackage{newpxtext,newpxmath}

\usepackage{authblk}

\title{Towards Green Computing: A Survey of Performance and Energy Efficiency of Different Platforms using OpenCL}
\author[1,2]{Philip Heinisch}
\author[1,2]{Katharina Ostaszewski}
\author[1,3]{Hendrik Ranocha}
\affil[1]{Institut für angewandte numerische Wissenschaft e.V. (IANW), Braunschweig, Germany}
\affil[2]{Institut für Geophysik und extraterrestrische Physik (IGeP), Technische Universität Braunschweig, Germany}
\affil[3]{King Abdullah University of Science and Technology (KAUST), Thuwal, Saudi Arabia}
\date{March 8, 2020}
\makeatletter
\hypersetup{pdfauthor={Philip Heinisch, Katharina Ostaszewski, Hendrik Ranocha}}
\hypersetup{pdftitle={\@title}}
\makeatother

\begin{document}

\maketitle

\begin{abstract}
When considering different hardware platforms, not just the time-to-solution can be of importance but also the energy necessary to reach it. This is not only the case with battery powered and mobile devices but also with high-performance parallel cluster systems due to financial and practical limits on power consumption and cooling. Recent developments in hard- and software have given programmers the ability to run the same code on a range of different devices giving rise to the concept of heterogeneous computing. Many of these devices are optimized for certain types of applications. To showcase the differences and give a basic outlook on the applicability of different architectures for specific problems, the cross platform OpenCL framework was used to compare both time- and energy-to-solution. A large set of devices ranging from ARM processors to server CPUs and consumer and enterprise level GPUs has been used with different benchmarking testcases taken from applied research applications. While the results show the overall advantages of GPUs in terms of both runtime and energy efficiency compared to CPUs, ARM devices show potential for certain applications in massively parallel systems. This study also highlights how OpenCL enables the use of the same codebase on many different systems and hardware platforms without specific code adaptations.
\end{abstract}

\section{Introduction}
The maximum useful clock frequency of modern processors has become increasingly limited by signal propagation delays and power dissipation, the so-called power wall. To support the growing need for computational power, multi-processor architectures have become the new de-facto standard in consumer, industrial and scientific applications. Together with the mainstream adoption of multi-core CPUs, initially task-specific Graphics Processing Units (GPUs) developed into highly parallel General-Purpose computing on Graphics Processing Units. Simultaneously, ARM-based RISC processors transformed from application specific embedded processors to low-cost alternatives to the CISC-based x86 processors, driven by the fast increase in computational requirements for mobile and embedded devices. In recent years, accelerators using Field Programmable Gate Arrays (FPGAs) or Application-Specific Integrated Circuits (ASICs) have also become available, e.g. to solve specific proof-of-work based system as used in many cryptocurrencies.

In the past, most applications were developed for a specific CPU architecture with little need for portability. While the x86 architecture was the only option for computationally intensive problems, these processors were unsuitable for mobile applications, due to the high power consumption and need for external support circuitry. Low-power ARM-based processors could easily be integrated into battery powered mobile devices but lacked the computational resources (i.e. dedicated floating point units) and support for large amounts of RAM or high-speed data buses needed for complex data processing or scientific computations. The use of FPGAs, ASICs and GPUs was limited by the high cost of soft- and hardware development.

This has changed drastically in recent years, with increasingly power efficient x86 hardware, powerful multi-core ARM-based System-on-a-Chip (SoC) designs and the widespread adoption of low-cost high-powered GPUs both as external hardware and integrated with ARM and x86 CPUs. Even FPGAs became available as accelerator cards or integrated within CPUs (e.g. Intel Xeon Gold 6138P or Xilinx Zynq). Combined with the increasing support for portability between different architectures by software development kits, libraries, and compilers this lead to the concept of heterogeneous computing. Frameworks like Khronos' Open Computing Language (OpenCL), Microsoft's C++ AMP or higher level programming models like Intel's oneAPI or SYCL as a higher-level programming model for OpenCL provide the necessary software portability to target these heterogeneous platforms with a single code-base, but still with limitations regarding performance portability \citep{Agosta2014,Pennycook2013}. In many studies, the question of performance portability is primarily discussed in relation to runtime, but the problem also arises in regard to power consumption. Limiting factors for portability for example with OpenCL are plentiful and can range from workload sharing between compute units or cores to the optimization and usage of specific hardware acceleration features. Especially for embedded or mobile devices, efficient usage of specific hardware features can have a large impact on the required power, while having comparatively little impact on overall runtime.

With all these hardware options readily available, the question arises which combination of devices is best suited to solve a given set of problems. The answer depends not only on the time-to-solution but also on the energy-to-solution and the price of the hardware. Especially the energy necessary to reach a solution becomes increasingly important as battery powered systems have to handle computational intensive tasks like image processing for autonomous electric vehicles or machine learning. Even for large-scale data centers or high performance computing (HPC) facilities the energy cost over the lifetime of the systems can be higher than the initial cost of acquisition. This has lead to programs like the Green500 list \citep{Sharma2006} to rank supercomputers based on energy efficiency. In this context the term performance per watt is widely used. This metric is of course linked to accompanying benchmark workloads. To provide some comparability in many cases the linear algebra LINPACK suite is used \cite{Dongarra}, which primarily relies on standard microbenchmarks. This study aims at using complete self contained real world examples from computational physics and mathematics as well as signal and image processing to show the portability of a given implementation without specific optimizations and the differences in time- and energy-to-solution for different platforms. Due to its versatility and support by many different manufactures, OpenCL was chosen to execute the set of testcases. Hence, this study can not only help answer the question of which kind of hardware is best suited for a given problem, but it also investigates the capabilities of OpenCL on a variety of hardware platforms.

As the results will show, different devices are optimized for certain workloads or are even missing certain hardware features like support for double or half precision floating point arithmetic. Especially with more task specific devices, calculating the performance per watt based on a specific benchmark testcase (like LINPACK) is not representative for a different set of tasks and no general performance per watt exists. While such a measure can be used for more abstract comparisons or theoretical benchmarks, it is of less use when designing real world systems. Hence, as part of this work several different testcases were used and the results will be called energy-to-solution in the following, to clarify that these values are always relative to a specific problem set and cannot necessarily be generalized.

Previous studies have instigated the advantages of different architectures related to time-to-solution mostly in the context of high-performance or scientific computing linked to very specific applications like Monte Carlo methods \citep{Weber2011}, solvers for systems of linear equations \citep{CHEIKAHAMED201732}, aerodynamic Navier--Stokes solvers \citep{Pennycook2013}, or mesh interpolation \citep{BUYUKKECECI201394}. Research on the trade-offs between time-to-solution and energy-to-solution has mainly focused on the comparison between ARM-based processors and x86 CPUs in the context of computing clusters \citep{GoDdeke2013,Padoin2015,Bez2016} or was focused specifically on GPUs \citep{holm2019gpu}. To measure the power requirements, the integrated power profiling features available in modern CPUs and GPUs were used. Knowledge of the realtime power consumption of compute devices is not only important to the user to optimize code efficiency but also an integral part of automatic power state and dynamic clock frequency management. Hence, manufacturers have started to implement both software based heuristics and hardware based current and voltage monitoring to provide power estimates for different components with very little overhead. These vendor specific profiling solutions have a relatively high temporal resolutions, typically with maximum sampling rates in the range of 100 Hz and can be accessed by the user through vendors specific APIs.

All test-cases were implemented using OpenCL, as it is not only promising for scientific HPC \citep{Ostaszewski2018}, but it also allows to target heterogeneous systems even for mobile applications \citep{VALERY201844}. OpenCL also has the major advantage of transparently abstracting the low-level parallelization from the user. Targeting both CPUs and GPUs directly with C/C++ code on different operating systems would otherwise require different implementations, as only OpenCL allows targeting heterogeneous systems running one of the major operating systems. Additionally this opens up the possibility to investigate the performance portability of typical algorithms, which becomes an increasingly important question with the adoption of GPUs \citep{holm2019gpu} and ARM-based systems both in scientific HPC and in consumer applications.

\section{Methods and Test Cases}
\label{methods}
Four different test cases were chosen and implemented in OpenCL to test both the time-to-solution and energy-to-solution on different platforms. The open source cross platform tool ToolkitICL \citep{heinisch2019toolkitICL} was used as a framework on the host to execute and profile the different kernels. This tool was specifically designed to execute generic OpenCL kernels on HPC systems both for production use and for profiling and benchmarking. The data input and output, kernel source, and OpenCL settings are completely handled by HDF5 files. This makes it possible to encapsulate the individual test cases into separate HDF5 files and use the same host application for all runs to ensure comparability.

ToolkitICL also includes support for power and temperature profiling for AMD, Intel, and Nvidia hardware. While the sampling rates (typically around 10 Hz) achieved by these profiling solutions do not allow for instruction level power profiling (see e.g. \citealp{Mukhanov:2017:AFE:3058793.3050436}), they are sufficient to estimate the total energy necessary to execute a larger set of instructions. To gauge the accuracy of these built-in power profiling tools, a true RMS digital multimeter was used to log the current consumption during execution on different platforms to provide a measurement based reference. The conversion efficiency of the power supply (typically between 80\% and 90\%) was also considered and the measurements corrected based on the specific device to represent the actual power consumption. All test runs were done with and without power profiling enabled, to gauge the overhead associated with software based profiling. To determine the energy-to-solution, the time series of power measurements (either from profiling API or current measurement) was numerically integrated over the OpenCL kernel execution time. To prevent bias caused for example by storage or networking hardware, the baseline idle power consumption was determined and subtracted from the power measured during execution. Each run was repeated at least three times and only the averaged results were used for the final analysis.

Due to the large differences in architecture between hardware platforms, selecting generic test cases without intrinsically favoring specific setups is a challenge. Especially while testing highly parallel applications on multi-node cluster setups, like the one used by \cite{GoDdeke2013}, performance can be dominated by the efficiency of interconnect networks and memory architecture. The results are therefore only meaningful for a specific system and care must be taken when these results are generalized. Hence for this study benchmarking was only performed on single nodes to eliminate the influence of network architecture and additional overhead associated with distributing the workload across nodes. To avoid the caveats of classical microbenchmarking while still being able to use less powerful ARM systems incapable of running for example large numerical simulations, a compromise between reducing the complexity of the program (i.e. multi-node support or file in- and output) and reducing the size of the problem itself (i.e. number of timesteps or data points) was chosen. OpenCL was selected to realize the test cases not only because it can target all intended platforms using the same source code but also because it has execution time profiling features already build in. This is especially important when comparing accelerators (i.e. GPUs) with dedicated memory against CPUs, as the additional memory transfer overhead has to be taken into account. To simulate extensive real world workloads, the individual test cases were executed back-to-back several thousand times using the same input data, to yield an average time- and energy-to-solution.

OpenCL uses vendor specific platform driver implementations to compile and run the OpenCL code on the specific devices. Especially for GPUs, the OpenCL implementation is provided directly by the manufacturer as part of the GPU driver. Currently, only Intel supplies OpenCL CPU drivers for their devices. AMD discontinued CPU based drivers and ARM only provides OpenCL GPU drivers. While optimized for their own set of CPUs, the Intel platform drivers can be used on AMD devices, but the performance is not guaranteed. Another open-source cross platform OpenCL implementation is the "Portable Computing Language" (POCL) \cite{Jaaskelainen:2015:PPO:2812638.2812652}. It can be compiled directly on the target system, to achieve the best possible optimization for the specific hardware. POCL was used in this study for all ARM CPUs and tested on some of the Intel devices for reference. For AMD CPUs, POCL, Intel and the legacy AMD drivers were tested.

\subsection{OpenCL Test-Cases}
Different computational problems were selected for testing to represent a wide variety of real world workloads encountered in scientific and commercial applications. Testcases were chosen specifically to represent not just microbenchmarks, but rather a selection of real-world problems comprised of more abstract benchmark cases and completely self contained filter or simulation implementations. The OpenCL benchmarks were tested against similar implementations of the algorithms in MATLAB, CUDA, C++ AMP, and C++/OpenMP to verify the results and ensure that comparable or better time-to-solution was achieved by the OpenCL implementation. This approach was also used to ensure that none of the testcases intrinsically favours one architecture. While it has been  All necessary testcase files are open source and available as part of the ToolkitICL tool \citep{heinisch2019toolkitICL}.

\subsubsection{2D Median Filter}
\label{2dmedian}
The median filter was selected because it is a common nonlinear edge preserving digital filtering technique typically used to reduce image noise \citep{Lim1990}. While it is also used in 1D signal processing, it is most often applied as a stand alone filter in image editing or as a pre-processing step in image recognition or classification applications. As such it is used in real time image processing systems, which are often constrained by power limitations. It is also a typical candidate for a problem theoretically benefiting from sharing workloads between a CPU acquiring the input and an accelerator like an embedded GPU to handle the actual processing. The idea behind median filtering is to replace each pixel with the median of a window of n$\times$n neighboring pixels. As image processing is mostly done on integer datatypes, this example was implemented solely using integer operations and does not use floating point variables. A 4K (3840×2160) image was used and 4000 frames processed to avoid single execution artifacts and to provide a realistic workload. As the underlying algorithm can easily be parallelized and implemented independently of hardware specifics, this example was also used to verify the OpenCL runtimes with a pure CPU based OpenMP \citep{dagum1998openmp} implementation. As with the OpenCL implementation, no specific optimization were used and the algorithm was implemented nearly identical to the OpenCL version.

\subsubsection{Dot Product}
The dot product is a standard algebraic operation, calculating the sum of the products of the corresponding entries of two vectors of equal length. As a typical microbenchmark, even though it is a more abstract test scenario it combines a large number of multiply–accumulate operations, which can be used as a measure of the overall performance of a computational system. Additionally, the dot product reduces the input data to a single output value, which is inherently difficult to parallelize, as the individual steps are not independent. This kind of reduction requires atomic memory operation which can incur large performance penalties, especially on GPUs. Hence, the dot product can be calculated more efficiently using a single threaded approach (in particular by employing techniques like loop unrolling), even for relatively large vectors, but as many different algorithms in computational mathematics and data processing require some kind of atomic data reduction, this dot product example can be considered as a generic testcase for these kinds of reduction-requiring algorithms.

\subsubsection{Cross--Correlation}
Cross-correlation is a signal processing technique used to measure the similarity between two datasets depending on the displacement of one of the inputs. It is typically used for pattern recognition, machine learning, for example 1D speech recognition or 2D image pattern matching (e.g. \cite{Benesty2009}) or to find statistical links between datasets for example in climate analysis. Cross-correlation also has applications in large scale scientific data processing for time of flight or propagation analysis (e.g. \citep{PHMNRAS}). In these scientific applications, cross-correlation analysis is applied to extremely large data sets, requiring significant computational resources. Discrete implementations typically shift one of the signals relative to the other and calculate the correlation coefficient for each shift. As this workload can be separated into independent steps, cross-correlation can significantly benefit from parallel execution, especially on real time systems. This example implements a complete correlation analysis but with reduced dataset length using single precision floating point variables, which is typical for many real-world applications to conserve memory. 

\subsubsection{Runge--Kutta Differential Equation Solvers}
\label{RK}
Numerically solving (partial) differential equations is a problem encountered in many fields not only computational mathematics, but also physics, engineering and economics. Runge–Kutta methods are a set of iterative algorithms, widely considered as one of the de-facto standards to numerically approximate the solution of these equations. This testcase was not just implemented as a microbenchmark of a specific solver, but as full 2D simulation running for several thousand timesteps. It is also well-suited to compare the influence of the floating point precision on the time- and energy-to-solution, by executing the same algorithm using both single precision and double precision floating point variables. An open-source implementation with a corresponding physical problem used in current mathematical and physics research was chosen \cite{ranocha2018numerical,ranocha2018induction}, as the intention of this work is to use benchmark testcases as close to real world applications as possible.

\subsection{Hardware}
The hardware used for this study was selected to represent different architectures and vendors while focusing on typical platforms widely available on the market and in use in scientific HPC. It was not limited to higher-end enterprise level devices, as applications like image processing are required to run even on lower-end systems. Linux, Windows, and Mac OS X operating systems were used to exclude a possible bias due to differences in the operating systems. In total, 22 different discrete GPUs, three different integrated GPUs, and 14 server and workstation CPUs manufactured by Intel, AMD and Nvidia were tested. As ARM-based system are becoming increasingly important, four ARM-based CPUs and two integrated ARM GPUs were tested as well. A complete list of devices and operating systems is available as supplementary material. No specific performance related changes, like overclocking, were made to the systems.

\section{Results}
The computational power of the used devices is vastly different. Therefore, the time-to-solution is not directly comparable. In contrast, the energy required to solve the given problems is not directly related to the computational power of the device but rather to the energy efficiency or performance per watt. This of course assumes that only the power required to run the actual computational device is considered, excluding peripherals, storage systems and other related hardware. To guarantee the validity of the results, it was ensured that each of the individual runs was within 15~\% of the average both for time- and energy-to-solution.
\subsection{Power Measurement}
\begin{figure}
\centering
\includegraphics[width=0.35\textwidth]{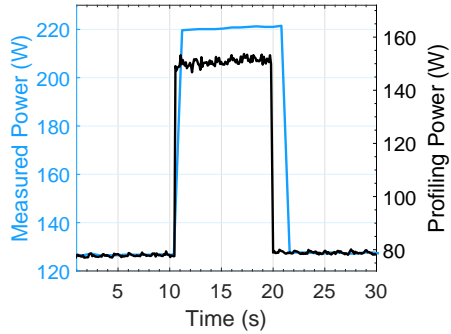}
\caption{Power requirements of a Tesla C2075 GPU during execution of the single precision Runge--Kutta differential equation testcase, determined by external primary current measurement (blue) and internal vendor specific power profiling (black).}
\label{fig:PWR}
\end{figure}

Fig. \ref{fig:PWR} shows the power requirements of an Nvidia C2075 GPU during execution of the single precision Runge--Kutta differential equation test case (see Section~\ref{RK}) over time. This example shows the results of both the internal vendor specific power profiling and external primary current measurement. As explained in Section~\ref{methods}, the externally measured power can vary from the internal profiling measurements. As overhead due to networking, storage, cooling or the host CPU which is not considered by the internal profiling solutions and can not fully be removed by subtracting the idle current. Furthermore, external measurements also track changes in power due to interconnect link state changes and other automatic workload dependent power saving features. The power supply efficiency ($\sim$90\%) was also not considered for this figure, to illustrate the difference. Additionally, the high-capacitance smoothing and filtering capacitors present in modern switch-mode voltage converters cause the slower rise and subsequently part of the slower decay in power after the computational task was completed. These capacitors are also the primary reason why the internal profiling measurements show higher frequency variations, which get filtered out and are therefore not present in the external measurements. This is not the case with internal profiling, as it either uses direct internal measurements of the computational device or a heuristic approach, which are completely independent of other peripherals or components. To minimize the influence of these external factors, which can significantly affect the comparability of the results, internal profiling was used for this study whenever possible. Only for older devices without power profiling support external measurements were performed an 10\% of the power subtracted to account for the average difference between internal profiling and external current measurement. The ARM devices lack support for power profiling as well, but as these self contained system-on-chip platforms are supplied by low voltage direct current (typically 5 V or 12 V) and lack high-speed interconnect buses or storage systems, external power measurements are more representative compared to traditional x86 systems.     
\subsection{Time-to-Solution}
\label{t-t-s}
\begin{figure}
\centering
\includegraphics[width=1\textwidth]{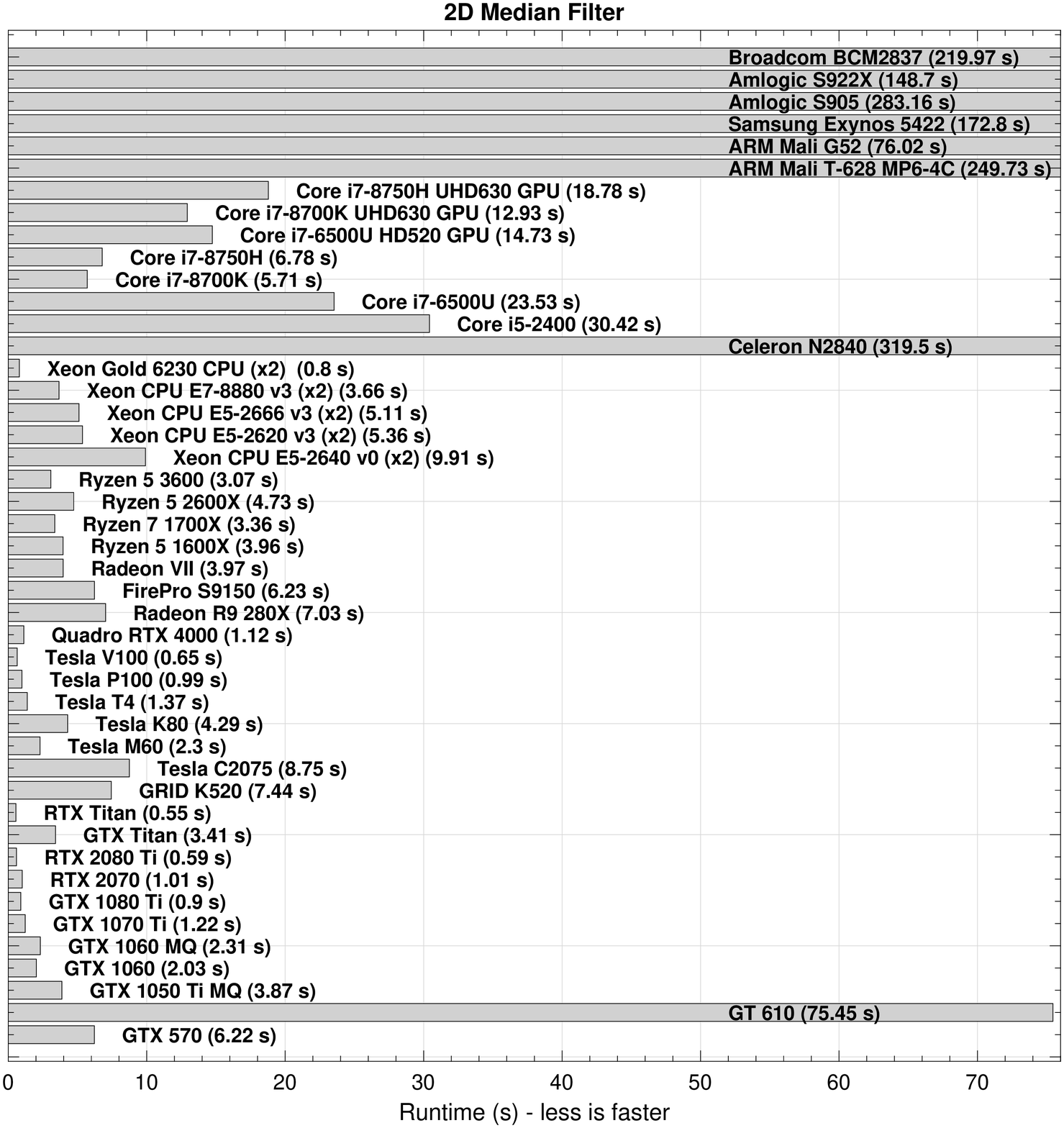}
\caption{Time-to-solution for the integer based 2D median image filter testcase for CPUs, GPUs and ARM devices.}
\label{fig:IMG_T}
\end{figure}

\begin{figure}
\centering
\includegraphics[width=1\textwidth]{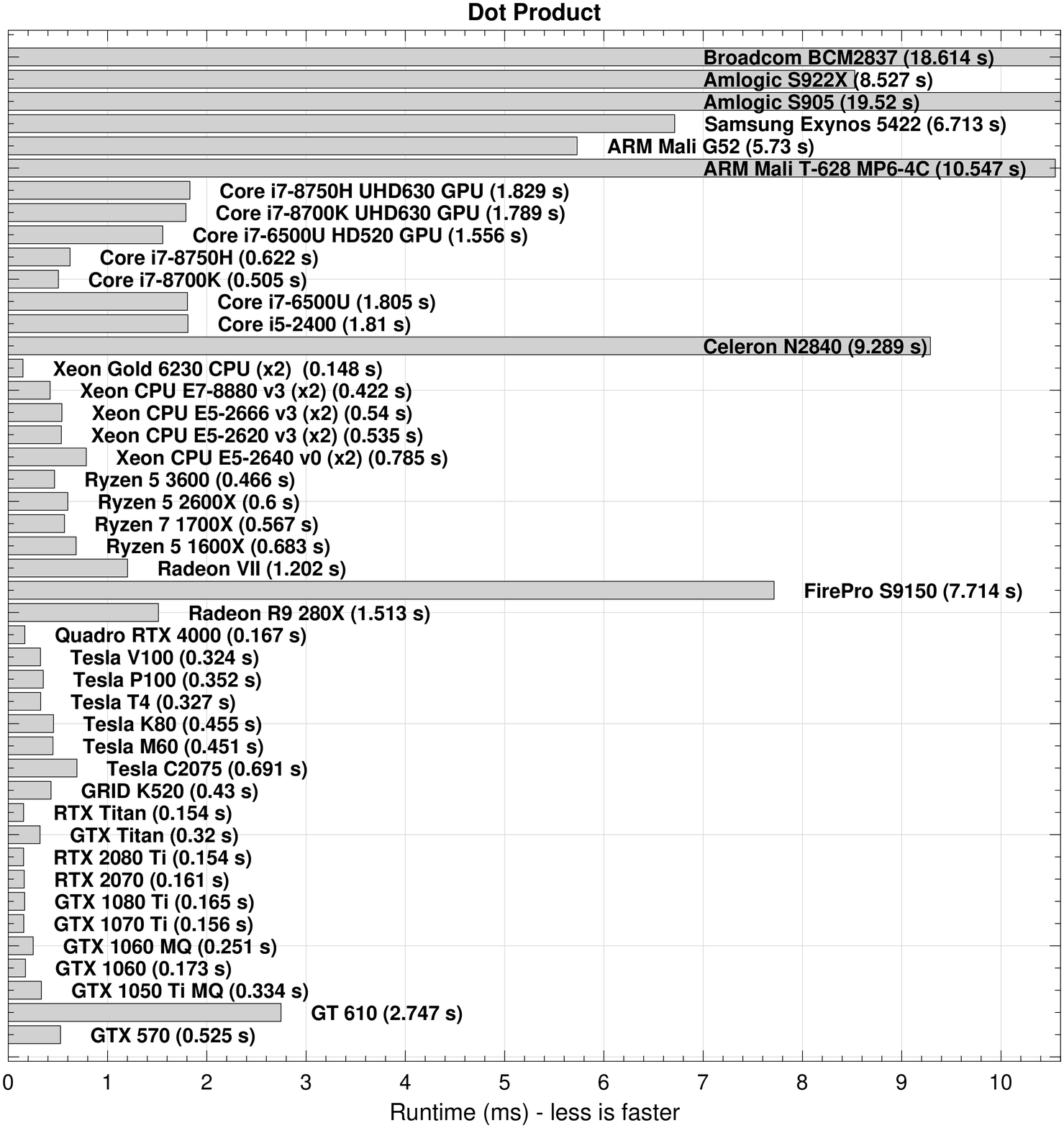}
\caption{Time-to-solution for the single-precision floating point dot product testcase for CPUs, GPUs and ARM devices.}
\label{fig:DP_T}
\end{figure}

\begin{figure}
\centering
\includegraphics[width=1\textwidth]{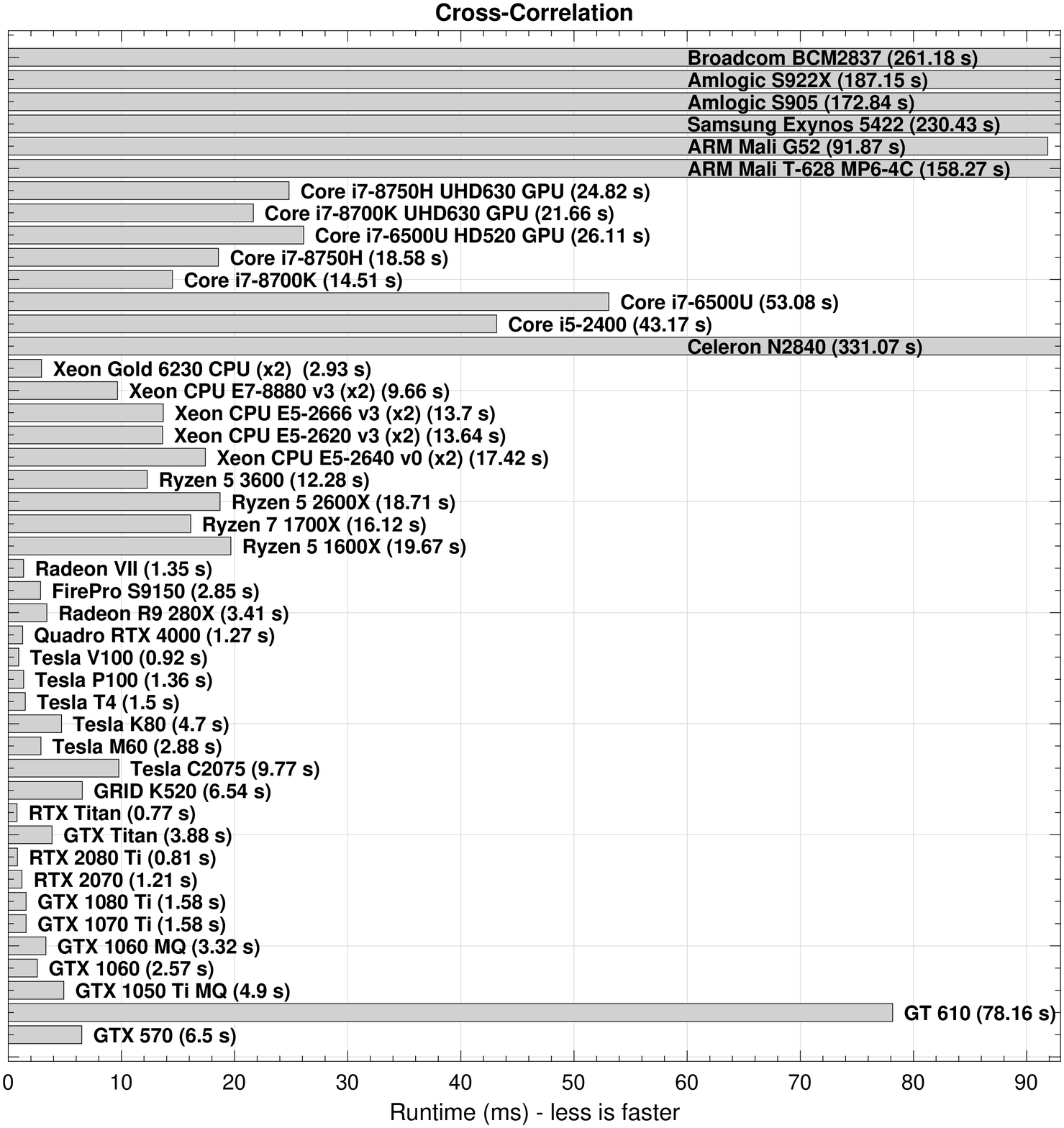}
\caption{Time-to-solution for the single-precision floating point cross-correlation testcase for CPUs, GPUs and ARM devices.}
\label{fig:COR_T}
\end{figure}

\begin{figure}
\includegraphics[width=1\textwidth]{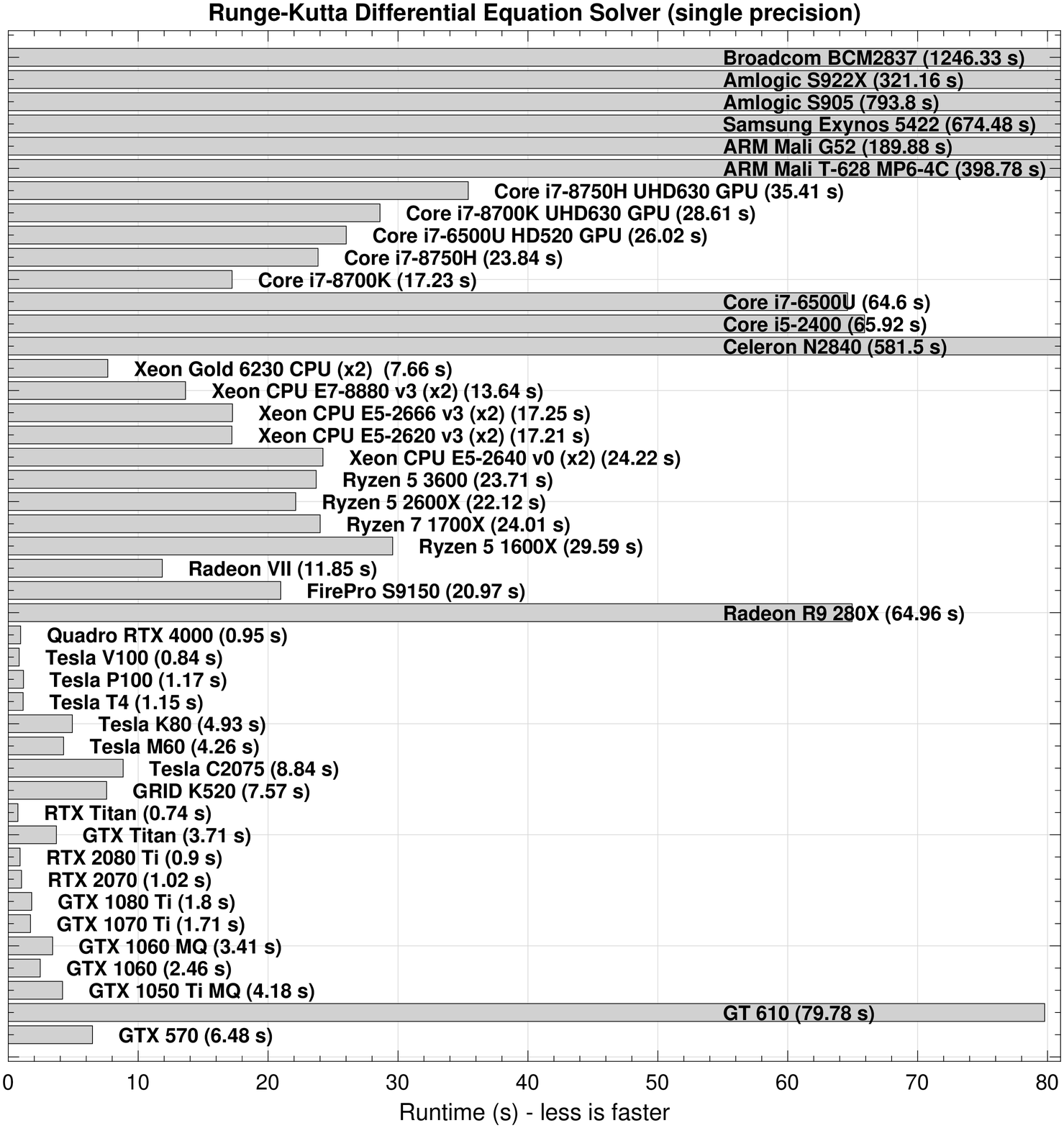}
\caption{Time-to-solution for the single-precision Runge-Kutta differential equation solver testcase for CPUs, GPUs and ARM devices.}
\label{fig:SSPRK_T}
\end{figure}

\begin{figure}
\centering
\includegraphics[width=1\textwidth]{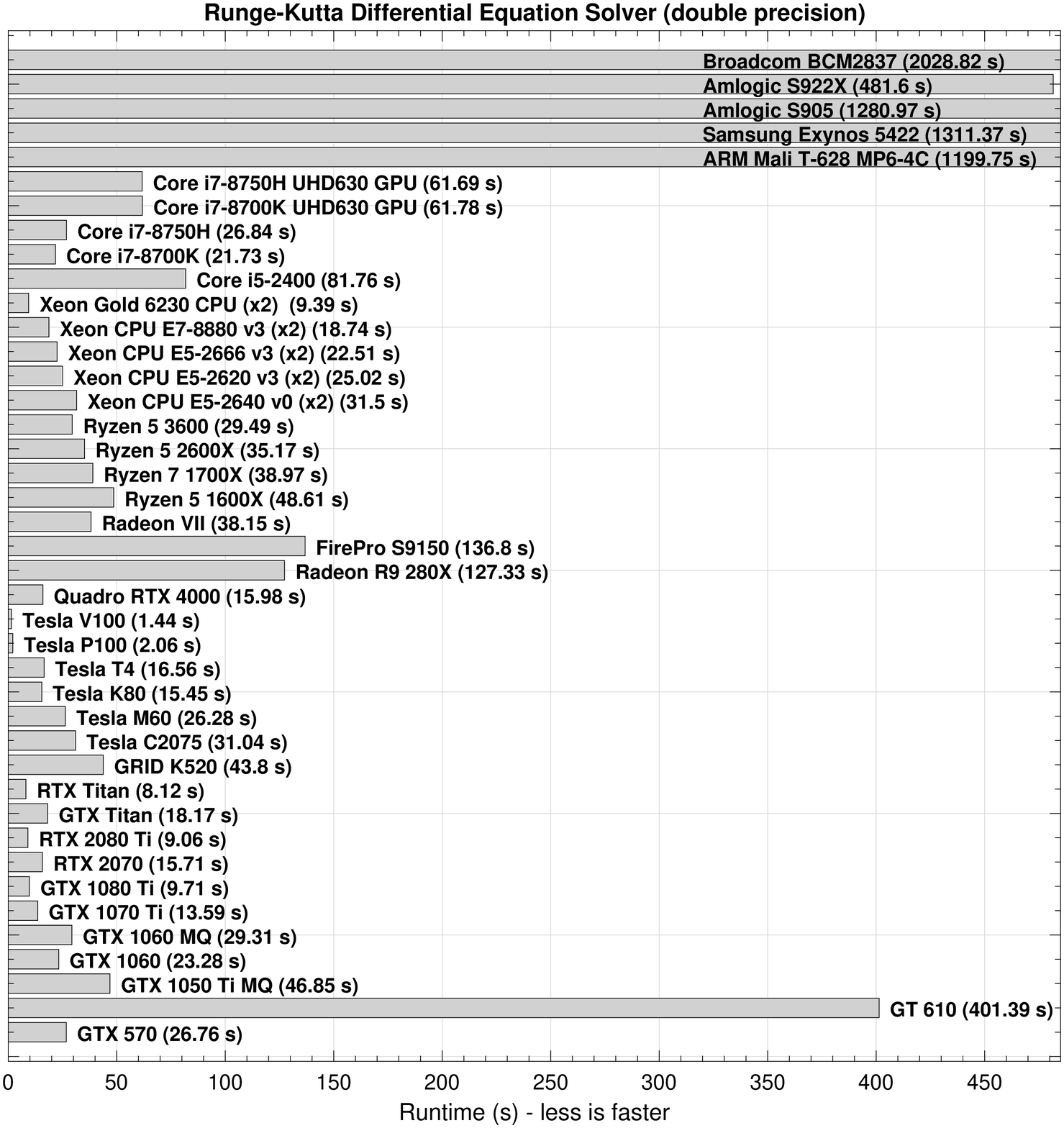}
\caption{Time-to-solution for the double-precision Runge-Kutta differential equation solver testcase for CPUs, GPUs and ARM devices.}
\label{fig:DSSPRK_T}
\end{figure}

Figures \ref{fig:IMG_T}--\ref{fig:DSSPRK_T} illustrate the execution time for the different benchmark test-cases for selected devices. Overall both energy- and time-to-solution improve with newer hardware. The latest GPUs performed best regarding time to solution, even outperforming the latest series of high-performance CPUs, especially with complex floating-point workloads. While older integrated GPUs are comparable in runtime to the actual CPU (like the Core i7-6500U), the CPU is noticeably more powerful in newer devices. Especially for the image processing example (see Fig. \ref{fig:IMG_T}), which is a prime candidate for shared memory based workload sharing between a CPU for acquiring and later evaluating the images and an integrated GPU used for image processing and filtering, the CPUs outperform the GPU by a factor of $\sim$ 2. In a real world application, the CPU would have the additional overhead of managing and synchronising the GPU tasks, which makes offloading less attractive and more challenging for the programmer. Hence, at least for the workloads used as part of this study, simply offloading computationally intensive tasks from the CPU to the integrated GPU does not positively impact runtime.

The advantage in execution time with GPUs is especially pronounced in tasks which rely heavily on complex single precision floating point operations, like the Runge--Kutta or cross-correlation examples. The difference in execution time is much less pronounced in double precision workloads, which is due to the fact that even most modern GPUs, with the exception of dedicated general purpose computing GPUs (like the Tesla V100/P100), are in contrast to CPUs not optimized for double precision tasks. In particular computing GPUs optimized for machine learning and inference tasks, like the Tesla T4, lack the hardware acceleration for double precision workloads.

Modern desktop CPUs like the Ryzen 7 or the Intel Core i7-8700K (both released in 2017) have already achieved the computational performance of older server grade Xeon CPUs like the E5-2640 (released 2012) even though they have less cores and lower energy requirements. While it was to be expected that the ARM devices are not comparable in runtime to modern CPUs or GPUs, it is noteworthy that the latest ARM GPUs like the Mali G52 (released 2017) approach the performance of older low-end GPUs and CPUs like the GT 610 (released 2012) or the Intel Celeron N2840 (released 2014). The possible advantages of ARM-based devices become more obvious when considering energy-to-solution.

To ensure that the vendor specific power profiling implemented in the ToolkitICL framework has no negative impact on the time-to-solution, baseline runs without power profiling were performed. As expected, no significant difference in runtime was found for GPUs and even for CPUs, the deviations were around 10~\%, which is within the statistical margin of error of this study. Slower systems like older CPUs or the ARM devices have no built-in power profiling capabilities, hence the impact of the profiling with computationally less powerful devices could not be investigated, as external current measurement had to be used.

The 2D median image filtering testcase was also used to compare the OpenCL performance with OpenMP for validation (see section \ref{2dmedian}). On average, the OpenMP implementation was slower by a factor of $\sim$ 3, which showcases the efficiency of the OpenCL framework in automatically optimizing the code for parallel execution. 

One notable exception for GPU performance are the AMD devices, which performed much worse than expected based on the computational power given by the manufacturer. Additional testing showed that AMD GPUs seem to require hardware specific OpenCL code tuning to even achieve performance roughly comparable to other GPUs or even CPUs. Due to the limited support for other software frameworks, it was not possible to check whether these tuning requirements are specific to OpenCL.

\subsection{Energy-to-Solution}
\label{e-t-s}
\begin{figure}
\centering
\includegraphics[width=1\textwidth]{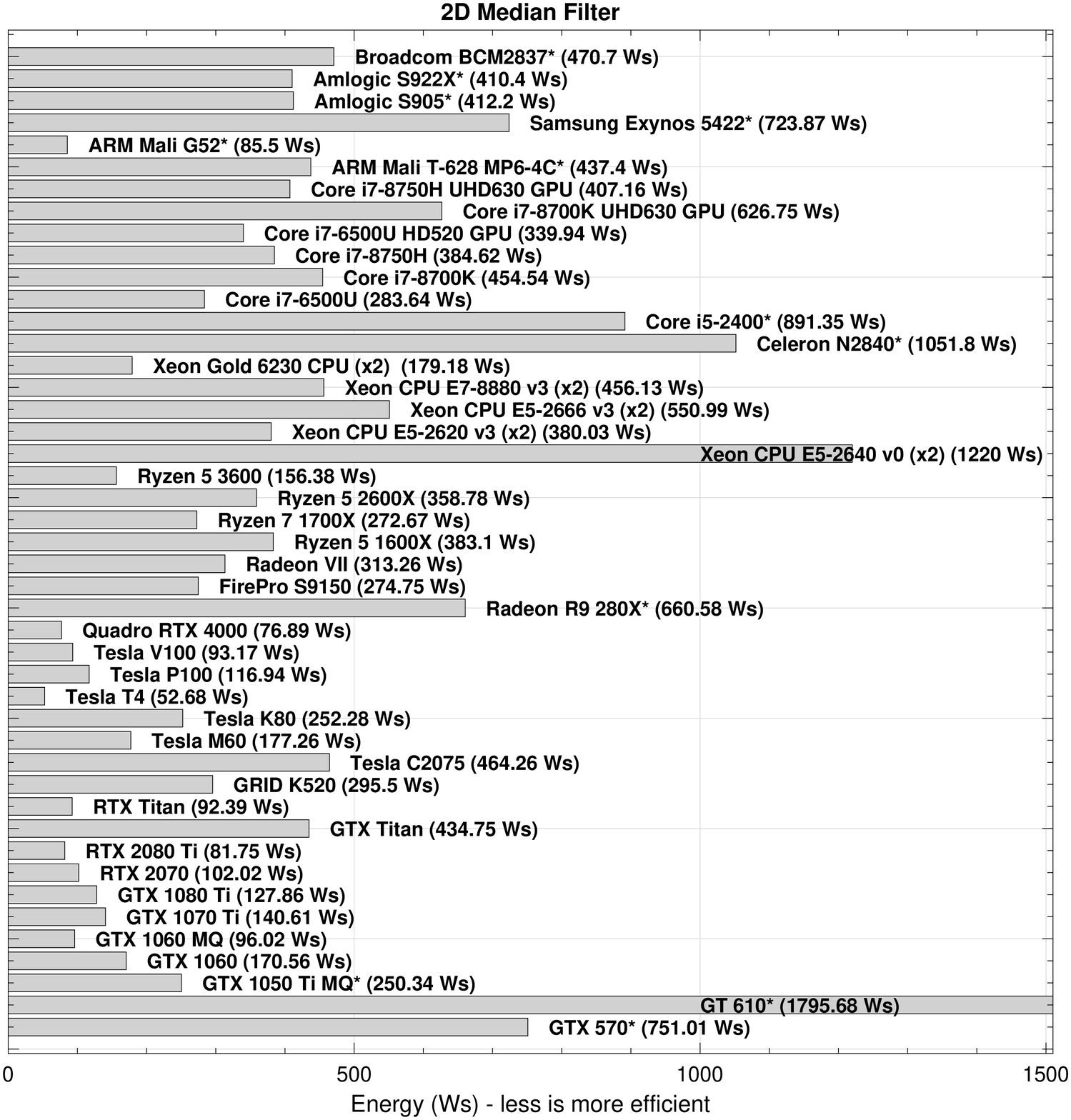}
\caption{Energy-to-solution for the integer based 2D median image filter testcase for CPUs, GPUs and ARM devices, results based on external current measurement are marked with an asterisk.}
\label{fig:IMG_E}
\end{figure}

\begin{figure}
\centering
\includegraphics[width=1\textwidth]{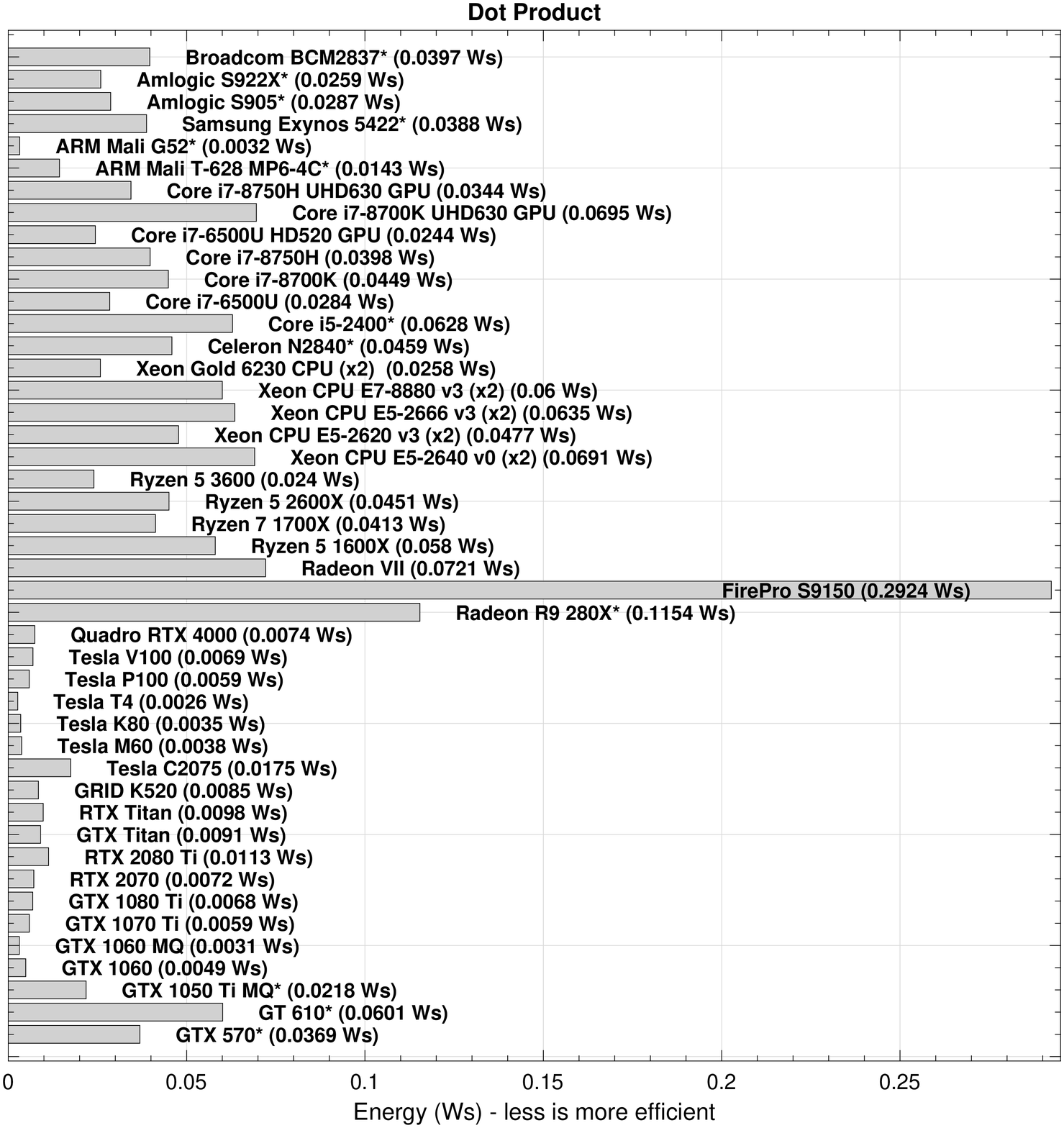}
\caption{Energy to-solution for the single-precision floating point dot product testcase for CPUs, GPUs and ARM devices, results based on external current measurement are marked with an asterisk.}
\label{fig:DP_E}
\end{figure}

\begin{figure}
\centering
\includegraphics[width=1\textwidth]{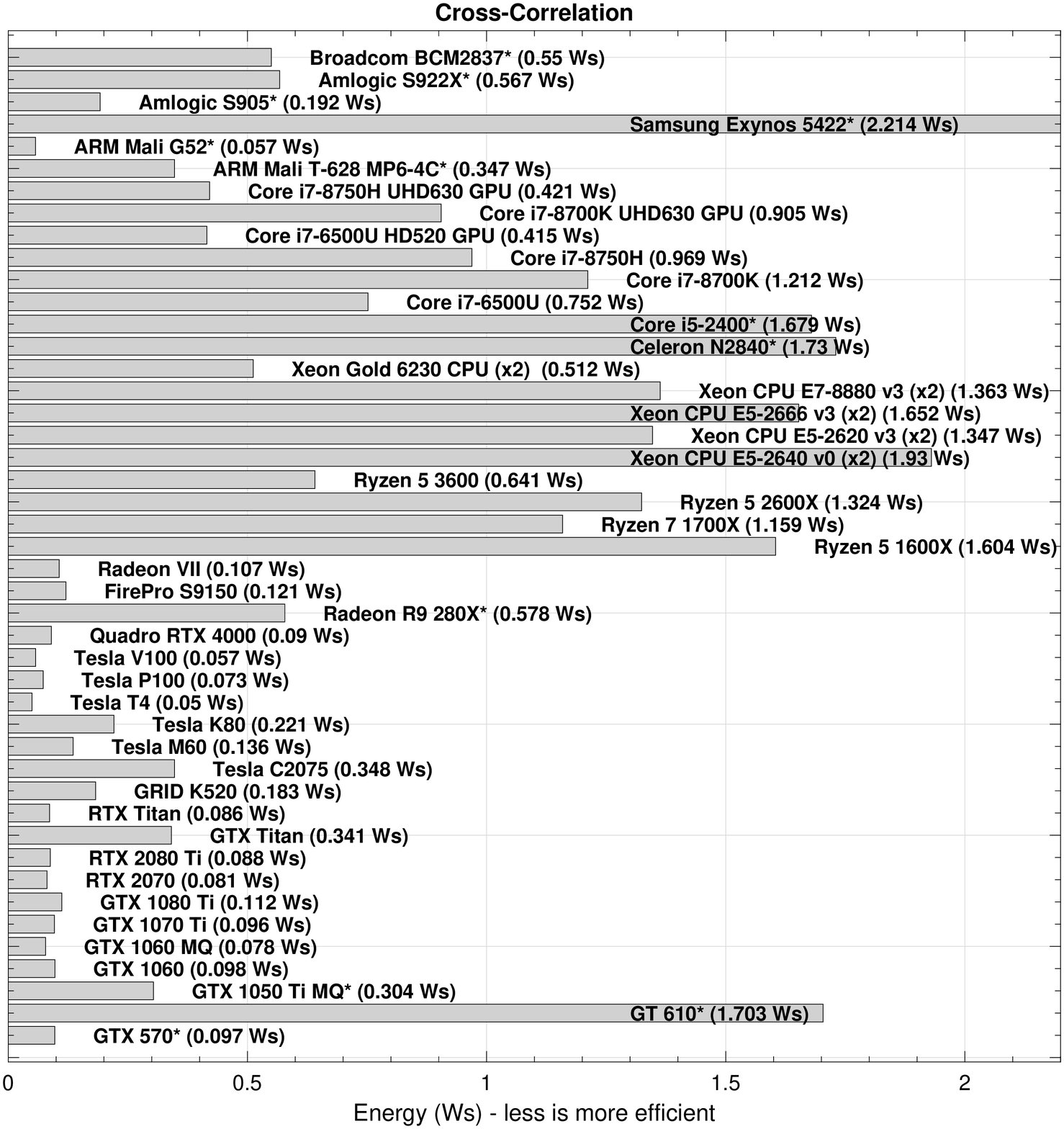}
\caption{Energy-to-solution for the single-precision floating point cross-correlation testcase for CPUs, GPUs and ARM devices, results based on external current measurement are marked with an asterisk.}
\label{fig:COR_E}
\end{figure}

\begin{figure}
\centering
\includegraphics[width=1\textwidth]{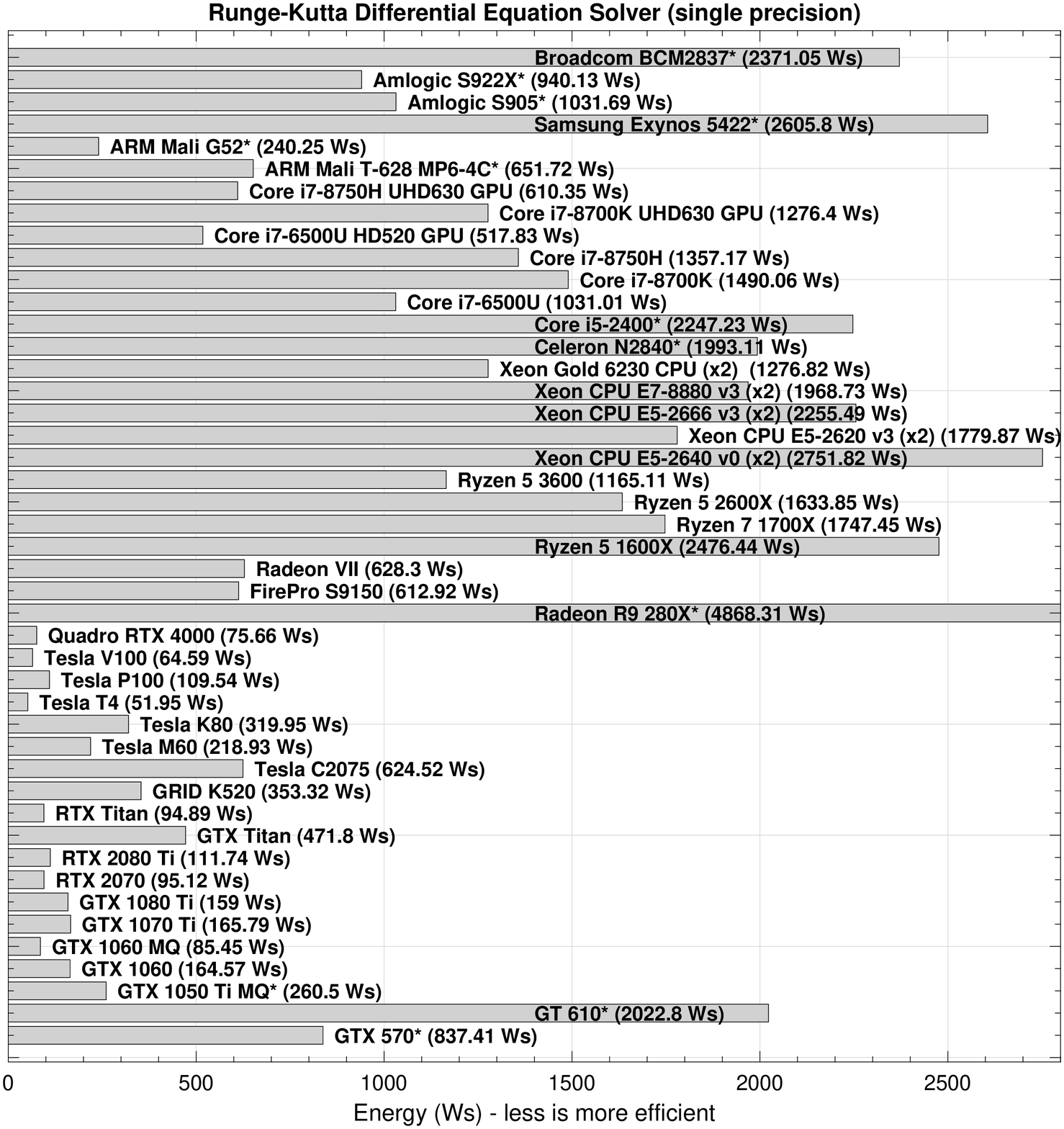}
\caption{Energy-to-solution for the single-precision Runge-Kutta differential equation solver testcase for CPUs, GPUs and ARM devices, results based on external current measurement are marked with an asterisk.}
\label{fig:SSPRK_E}
\end{figure}

\begin{figure}
\centering
\includegraphics[width=1\textwidth]{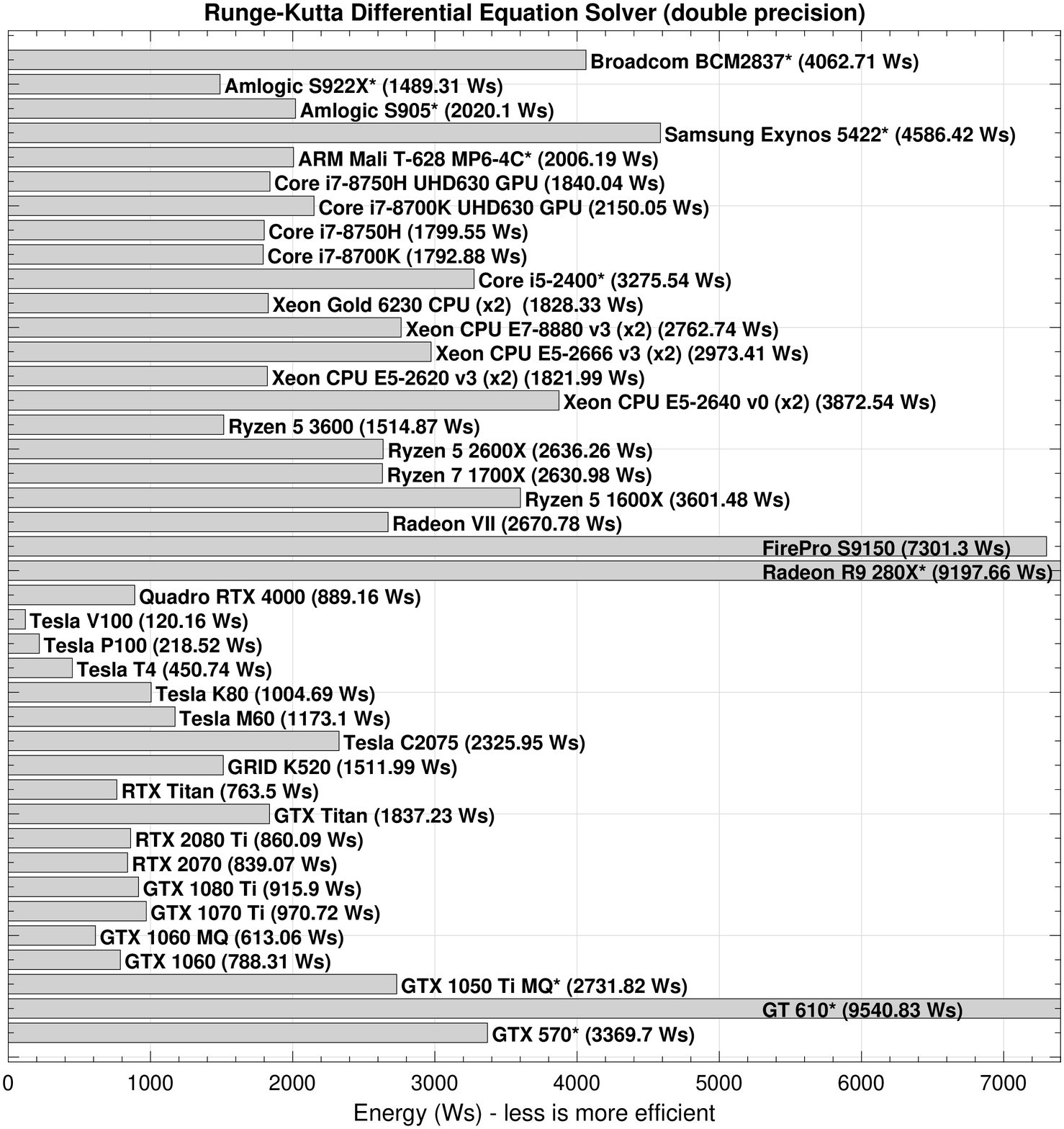}
\caption{Energy-to-solution for the double-precision Runge-Kutta differential equation solver testcase for CPUs, GPUs and ARM devices, results based on external current measurement are marked with an asterisk.}
\label{fig:DSSPRK_E}
\end{figure}

Similar to the runtime results, GPUs outperformed CPUs in energy-to-solution as well. The individual results are shown in Figures \ref{fig:IMG_E}--\ref{fig:DSSPRK_E}. As expected, CPUs targeted for mobile devices were more energy efficient compared to the much faster, but less energy efficient server CPUs. While the ARM devices were expectedly outperformed by traditional CPUs and GPUs in regard to runtime, the possible advantages of ARM-based systems become clear when considering energy-to-solution. In most cases both ARM CPUs and GPUs are comparable in energy efficiency to modern x86 CPUs and integrated GPUs. Especially with computationally simpler workloads like the dot-product example or signal processing workloads, modern ARM GPUs like the Mali G52 are even comparable or better then modern GPUs (see Fig. \ref{fig:DP_E}, \ref{fig:COR_E}). While they require longer to execute the tasks, ARM devices still require less energy overall. This exemplifies the potential of ARM devices as basis for energy efficient massively parallel compute clusters, for certain types of applications.

Integrated GPUs show no advantage over the corresponding CPU for integer and double precision floating point workloads. For single precision tasks, the integrated GPUs are up to a factor of two more efficient, but have comparatively minor trade offs in runtime (see Section \ref{t-t-s}). Especially with battery powered systems, offloading single precision tasks seems reasonable from an energy standpoint based on these results.

The power consumption for most of the devices (AMD, Intel and Nvidia) was measured using vendor specific internal power profiling APIs. As this approach is for many devices at least partially based on software heuristics and not just measurement, the accuracy of this method was validated for all APIs using external supply current measurement. For GPUs, the results were within 10~\%, while the deviation was below 15~\% for CPUs. The difference is most likely due to the fact, that for technical reasons, GPUs have at least some on-board power measurement, while most CPUs have not and rely completely on heuristics.

\subsection{Development over Time}
\begin{figure}
\centering
\includegraphics[width=1\textwidth]{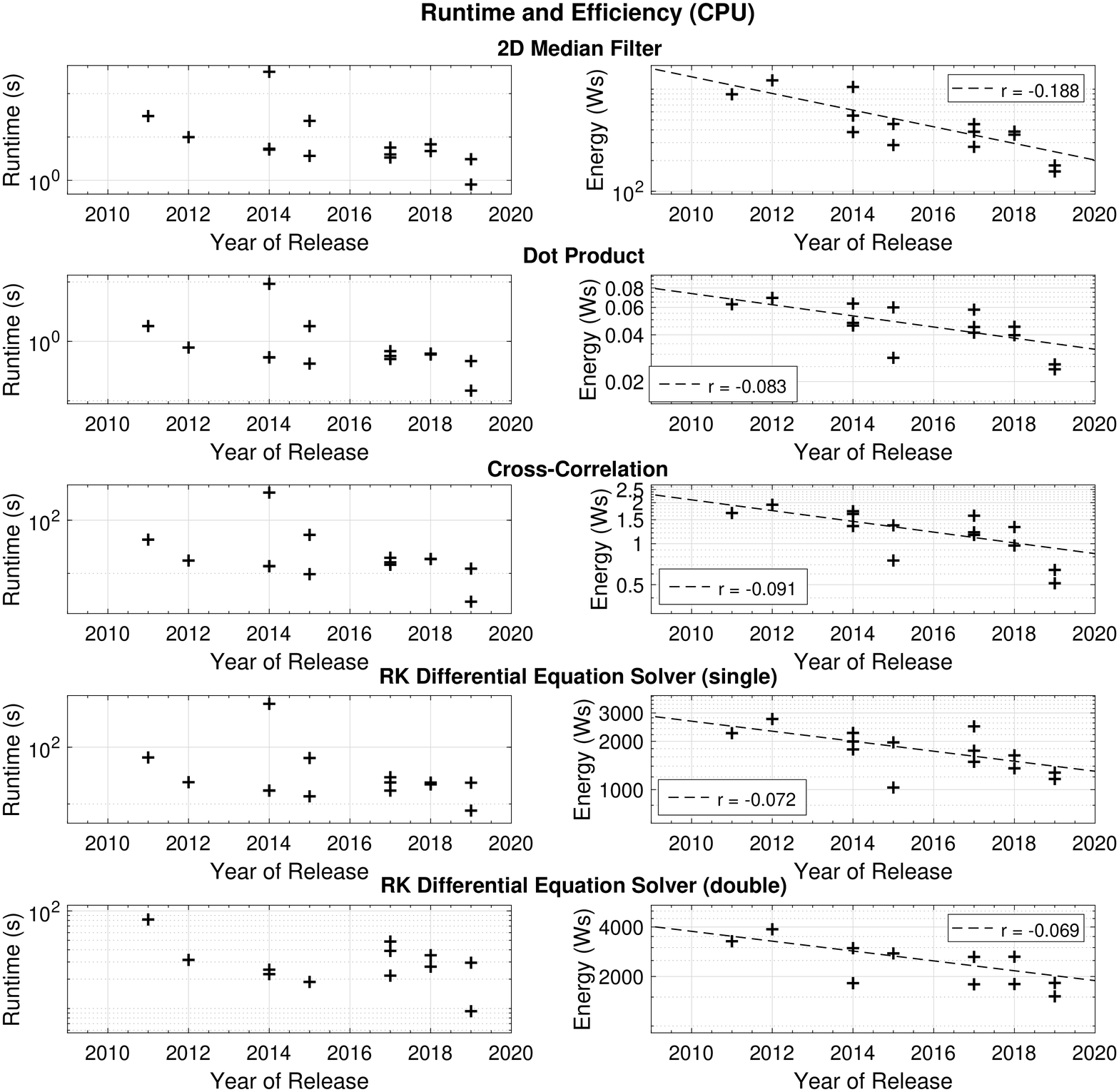}
\caption{Time-to-solution (left panels) and energy-to-solution (right panels) plotted logarithmically over the initial year of release of the device for all CPUs except ARM devices. An exponential fit was used to approximate the development of the energy efficiency over time, the growth rates are given in the individual plots.}
\label{fig:ML_CPU}
\end{figure}

\begin{figure}
\centering
\includegraphics[width=1\textwidth]{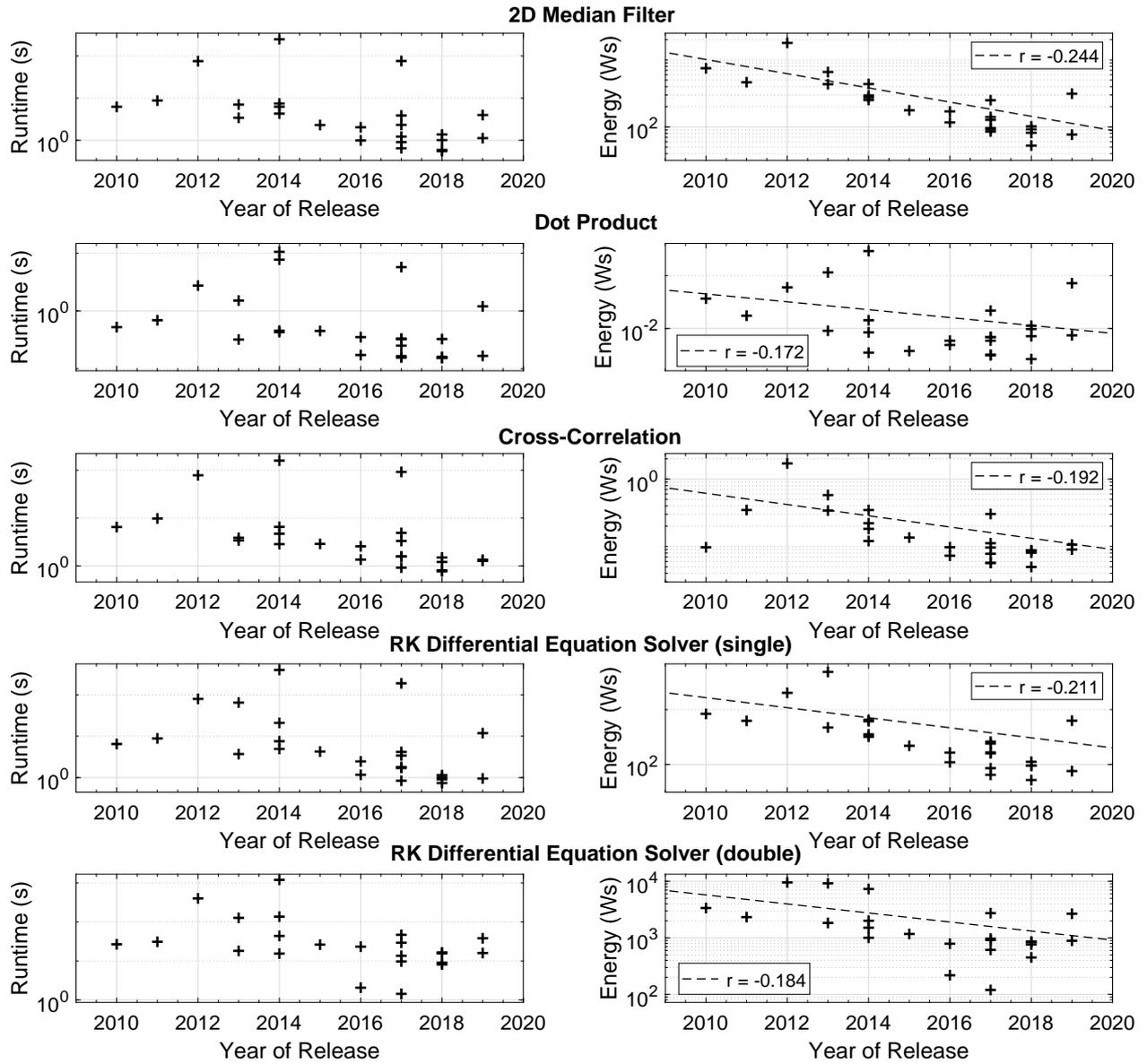}
\caption{Time-to-solution (left panels) and energy-to-solution (right panels) plotted logarithmically over the initial year of release of the device for all discrete GPUs. An exponential fit was used to approximate the development of the energy efficiency over time, the growth rates are given in the individual plots.}
\label{fig:ML_GPU}
\end{figure}

As benchmark results are available for devices released between 2010 and 2019 spanning almost an entire decade, the development of the time-to-solution and energy-to-solution for the different devices has been investigated. Compared to CPUs, GPUs were initially intended as task specific hardware accelerators and the development of GPUs into more general purpose computing devices is a rather recent trend. Hence, Fig. \ref{fig:ML_CPU} and Fig. \ref{fig:ML_GPU} show the results separately for CPUs and discrete GPUs. Based on the work of \cite{ML}, the results were plotted semi-logarithmically. The runtime, especially for CPUs, shows no clear trend over time, which can primarily be attributed to the fact that an intentionally broad selection of devices with different computational capabilities was selected. It is unsurprising that older enterprise level server CPUs are still faster than modern consumer or even embedded CPUs. The selection of GPUs is inherently more homogeneous, as it is primarily comprised of discrete GPUs, which are by design intended to be used for more computationally intensive tasks, otherwise integrated graphics solutions are used. Therefore, it is understandable, compared to the CPU results, that on average a decrease in runtime can be observed, but still without a statistically significant trend.

As already discussed in Section \ref{e-t-s}, the energy-to-solution or the related performance per watt is independent of the computational power. Increases in energy efficiency do not just translate into economical and ecological benefits, but also allow for more computationally powerful devices per unit area given a constant limit on thermal dissipation. Especially with modern many-core architectures, energy efficiency under load can be used as an overall metric of the computational capabilities of a system independent of the specific computational devices (i.e. number of cores or clock rates). With the set of devices used in this study, an exponential increase in energy efficiency over time was observed. An exponential function of the form $a \cdot \mathrm{e}^{r t}$, where $a$ is a problem specific scaling constant, $t$ is the year, and $r$ is the growth rate, was fitted to the results (see dashed lines in Fig. \ref{fig:ML_CPU} and Fig. \ref{fig:ML_GPU}). Based on these findings, on average the energy efficiency improves by a factor of 1.22 every two years for CPUs and by a factor of discrete 1.50 for GPUs, which is a difference of $\sim$ 19 \%. This shows the slow down in development relative to the doubling of the number of components per integrated circuit postulated by Moore's law \cite{ML}, which is linked to an even faster increase in performance per watt (based on the Dennard scaling law \cite{Dennard}). Even though the development seems to be slowing down, in general newer devices are still noticeably more energy efficient, but without comparable improvements in runtime. While this is less important for the average user, who is primarily interested in the time-to-solution, it significantly impacts the running costs of high-performance computing systems. 
\newpage
\section{Summary and Conclusion}
Using benchmarking testcases based on real world scientific and engineering workloads, this study showed that GPUs outperformed CPUs both in time- and in energy-to-solution. In particular with tasks relying heavily on complex single-precision floating point operations, even consumer grade GPUs can outperform server grade CPUs by a factor between 4 and 20 in runtime. While the difference in time-to-solution is less pronounced with double-precision or integer workloads (between a factor of 1.7 up to 10 on average speedup), particularly the energy efficiency of GPUs is superior and on average about an order of magnitude better, making GPUs a faster and more energy efficient alternative to CPU-only systems. On average the energy efficiency improves by a factor of 1.22 every two years for CPUs and by a factor of 1.50 for GPUs. This comparison is based on the actual execution time and the energy necessary to run the compute device without external peripherals, memory, networking or other components. While these factors play a role in the overall runtime and energy efficiency, these components are system specific, and therefore not comparable. While these factors can play a role in large cluster deployments, especially with extremely memory intensive operations, these factors become negligible with increasing interconnect bandwidth and GPU memory. While the energy requirements for storage and interconnect can be quite significant as well, they are nearly identical independent of the compute architecture. Although GPUs and other accelerators still need a host CPU, the impact on energy efficiency is comparatively low, as a single CPU can control multiple accelerators, while also sharing some of the compute workload by using heterogeneous compute frameworks like OpenCL. The measurements also showed that the CPUs required only between 10~\% to 15~\% of the maximum power under load during idle while just running the operating system and networking, so even if the CPUs in a cluster environment were just used to control accelerators like GPUs, the efficiency would still be favorable over a CPU only system. Integrated GPUs have significantly longer runtimes compared to the corresponding CPU (up to $\sim$ 3x) and are comparable in energy-to-solution for all tasks except single precision floating point, where they show an improvement of up to 2x. Even though integrated GPUs can offer significant advantages especially on mobile devices for specific tasks like video handling, these results show, that they offer no advantage over the actual CPU in more general purpose tasks like signal processing or numerical applications.

While single ARM devices are expectedly no match in runtime compared to CPUs or regular GPUs, the energy-to-solution or total necessary energy in particular for Mali GPUs is comparable or even better for single precision floating point workloads for example in signal processing. As no hardware specific tuning of the OpenCL code was performed for any of the tests it shows, that ARM-based devices are a reasonable alternative to x86 based systems in power constrained applications, even without hardware specific code adaptations, by making use of the OpenCL framework. It is noteworthy in this context that many of the ARM Mali GPUs lack native support for double precision floating point, which limits the use cases especially for scientific simulations. In the context of massively parallel computing clusters, these results show that up until now ARM-based systems are only a viable option for preferably single-precision workloads with very good strong scalability. The low overall power consumption combined with good energy efficiency allows a very large number of these devices to be clustered together creating a high core-count system with low per-core performance compared to the traditional higher per-core performance but comparatively lower core count x86 CPU based clusters.

As many high-performance computing clusters are required to solve a diverse set of problems in many instances without hardware specific code optimizations, the results presented in this study suggest that GPUs provide the most versatile solution. While achieving energy efficiency comparable to ARM-based devices, regular GPUs combine order of magnitude more computational power in a single device and outperform CPUs in time- and energy-to-solution. In particular machine learning optimized devices offer more economical solutions for integer or half/single precision floating point workloads. For applications relying heavily on double precision floating point calculations, only high end GPUs intended for general compute tasks offer significant improvements in runtime. Consumer GPUs and devices optimized for machine learning are comparable in runtime to CPUs, but lack the hardware necessary for fast double precision calculations. In comparisons, for small workloads especially on power constrained embedded devices, ARM-based systems provide clear advantages, as they are comparable or even better than x86 CPUs or even mid-range GPUs in energy efficiency under load. Especially as the integrated ARM GPUs share memory with the CPU cores, compute intensive tasks can efficiently be shared or even offloaded to the integrated GPU, reducing the memory copy penalty.

This study also illustrated how OpenCL can be used to deploy the same code to heterogeneous devices from different vendors spanning from regular CPUs to GPUs and even ARM devices without significant performance penalties or code optimizations. No adaptions to the code had to be made for the code to run on any of the devices and even though performance portability can still cause problems with OpenCL especially with some GPUs, overall no obvious outliers were detected, suggesting that with recent OpenCL implementations, even for reasonable complex programs, results are well within realistic expectations. This is especially useful, as the target devices for many applications both in scientific and commercial applications can change over the lifetime of the application, which required manual portation of the code or cross-compilation in the past.

As part of this study, the internal power profiling features of modern GPUs and CPUs were also validated both in terms of accuracy and associated overhead. The results showed that while not being good enough for instruction level profiling, sampling rates of around 10 Hz were possible and the software reported current was within 15\% of the values obtained by measuring the total system input current with an RMS multimeter.

\printbibliography

\end{document}